# A Study on the Performance of Generative Pre-trained Transformer (GPT) in Simulating Depressed Individuals on the Standardized Depressive Symptom Scale

Sijin Cai[1], Nanfeng Zhang[1], Jiaying Zhu[1], Yanjie Liu[1], Yongjin Zhou[1,2]*

[1] School of Biomedical Engineering, Health Science Center, Shenzhen University, Shenzhen, China
[2] Marshall Laboratory of Biomedical Engineering, Shenzhen University, China

*The corresponding author

**Corresponding author:**
 Yongjin Zhou: yjzhou@szu.edu.cn

**Abstract:**
**Background:** Depression is a common mental disorder with societal and economic burden. Current diagnosis relies on self-reports and assessment scales, which have reliability issues. Objective approaches are needed for diagnosing depression.

**Objective:** Evaluate the potential of GPT technology in diagnosing depression. Assess its ability to simulate individuals with depression and investigate the influence of depression scales.

**Methods:** Three depression-related assessment tools (HAMD-17, SDS, GDS-15) were used. Two experiments simulated GPT responses to normal individuals and individuals with depression. Compare GPT's responses with expected results, assess its understanding of depressive symptoms, and performance differences under different conditions.

**Results:** GPT's performance in depression assessment was evaluated. It aligned with scoring criteria for both individuals with depression and normal individuals. Some performance differences were observed based on depression severity. GPT performed better on scales with higher sensitivity.

**Conclusion:** GPT accurately simulates individuals with depression and normal individuals during depression-related assessments. Deviations occur when simulating different degrees of depression, limiting understanding of mild and moderate cases. GPT performs better on scales with higher sensitivity, indicating potential for developing more effective depression scales. GPT has important potential in depression assessment, supporting clinicians and patients.

## I. Introduction:
### 1.1 Current Status and Diagnosis Methods of Depression

Depression, as a common mental disorder, has been a burden for millions of people, including both young and elderly populations[1-3], and has imposed significant societal and economic pressure[3]. The main symptoms of depression include sadness, despair, loss of appetite, sleep disorders, and persistent fatigue[4]. Currently, the diagnosis of depression is usually based on the presence of the aforementioned symptoms for at least two weeks[5]. However, traditional diagnosis methods for depression mainly rely on self-reporting by patients and standardized assessment scales, which may introduce subjectivity and reliability issues[5]. The limited number of standardized assessment scales may cause patients to develop resistance towards repeatedly completing the same scale. In addition, the reliance on patients' self-reports in assessment scales makes the results susceptible to their subjective feelings and expressive limitations, thus affecting the accuracy of diagnosis. Given the limitations of traditional diagnosis methods for depression, the exploration of more objective and reliable methods has become increasingly urgent in recent years.

### 1.2 Development and Application of GPT Technology

Generative Pre-trained Transformers (GPT) technology is an artificial intelligence technique that utilizes deep neural networks to generate human-like responses in natural language processing tasks. Due to its capability of generating authentic, coherent, and contextually-relevant textual responses, GPT technology has been widely applied in the field of natural language processing. In recent years, researchers have begun to explore the potential application of GPT technology in the clinical field [6-9] and the field of psychology[10, 11], such as the diagnosis and adjuvant treatment of depression[12]. Depression scales are essentially pure language-based, and GPT, as a natural language processor, possesses corresponding advantages and potential. However, the application of GPT technology in the diagnosis of depression has not been validated. It remains to be verified whether GPT can comprehend mental health issues, diseases, and human psychological states, as well as possess a deep understanding of the thinking patterns and mental states of depression patients.

**1.3 Objectives of This Study**

The first objective of this study is to evaluate and determine whether GPT can accurately and objectively simulate depression patients and normal individuals, in terms of understanding their mental states based on depression scale scores. We will utilize standard depression scales as the measurement standards for depression symptoms and assess the performance of GPT in accurately and consistently reflecting depression symptoms.

The second objective of this study is to evaluate and determine whether GPT can accurately and objectively simulate individuals with different levels of depression based on depression scale scores. Specifically, our goal is to assess whether GPT can accurately and objectively understand the mental states of individuals with mild, moderate, and severe depression symptoms. We will use standard depression scales to measure depression symptoms in different groups and determine whether GPT can differentiate and accurately reflect the symptoms of individuals with mild, moderate, and severe depression.

The third objective of this study is to explore the influence of the granularity of depression scales on the simulation effect of GPT. Based on the previous two study objectives, we will investigate whether the scores obtained by GPT simulating individuals with different depression symptoms align with the scores specified by the scale for these individuals. If they align, it may indicate that the depression scale has a good effect on GPT simulation of individuals with depression, further suggesting that the analysis of depression symptoms using this scale is more reasonable.

**II. Methods**
**2.1 Selection of Depression Scales**

Three assessment tools related to depression were used in this study, including the Hamilton Rating Scale for Depression (HAMD-17)[13], the Self-Rating Depression Scale (SDS)[14, 15], and the Geriatric Depression Scale (GDS-15)[16, 17]. The selection of assessment tools was based on two main factors: 1) extensive use and recognition in clinical practice and research, and 2) representation of different types of depression assessment scales. When selecting SDS and GDS-15, we considered the fact that these tools are self-administered, thereby eliminating the need for individuals to complete assessments in hospitals or clinics. However, these two assessment tools differ in terms of sensitivity, with GDS-15 providing only a binary (yes/no) response level, while SDS uses a four-level rating system. When selecting HAMD-17, we considered its status as a standardized assessment tool frequently used by clinical physicians to diagnose and monitor depression. Overall, our selection of depression scales provides a comprehensive examination of the application of GPT in the field of depression assessment.

**2.2 Experimental Design**

Two independent experiments were conducted to design experiments for each selected assessment tool. In the first experiment, GPT simulated the response of normal individuals and depression patients when using the selected assessment tools. Specifically, GPT simulated thoughts and behaviors related to depression, such as feelings of despair and discomfort, and generated responses based on the corresponding assessment tools. Then, normal individuals were simulated to perform the same operation as a control group. In the second experiment, GPT simulated the responses of individuals with mild, moderate, and severe depression as well as normal individuals when completing the selected assessment tools. For the experiments involving SDS and GDS-15, GPT played different roles and answered questions based on the content of each scale, with the corresponding GPT responses recorded. For the experiments involving HAMD-17, GPT played different roles and simulated a scenario where a doctor assessed the corresponding GPT role based on the scale. GPT then evaluated the score given by the doctor for each item on the scale. In order to avoid any potential order effects as much as possible, all contents were cleared before each experiment started.

**2.3 Data Collection and Analysis**

To collect the required data for this study, response data generated by GPT in the first and second experiments were saved and analyzed. These response data were compared with the expected results of the GPT simulation, including those of normal individuals and patients with varying degrees of depression severity. By comparing the GPT responses to the expected results, we evaluated the degree to which GPT understands symptoms of depression. Statistical analysis was performed to explore whether there were significant differences in GPT performance under different experimental conditions and whether different depression scales affected GPT performance.

**III. Results**

**3.1 Performance of GPT as Normal and Depressed Participants**

The experimental results of GPT's performance in role-playing normal individuals and depressed patients were evaluated using three selected scales for depression assessment (HAMD-17, SDS, and GDS-15). In the experiment where GPT played the role of a depressed patient, the generated responses were completely consistent with the depression rating criteria of the corresponding assessment tools. In the experiment where GPT played the role of a normal individual, the responses generated were almost completely consistent with the normal individual rating criteria of the corresponding assessment tools, except for one result that differed from the rating standard when completing the HAMD-17. The results are shown in Table 1 and Figure 1.

Table 1: Experimental Results of the Performance of GPT as Normal Individuals and Depressed Patients When Completing the Three Selected Depression Assessment Tools

| Index | GDS-15 | SDS | HAMD-17 |
|---|---|---|---|
| sensitivity(%) | 100 | 100 | 100 |
| specificity(%) | 100 | 100 | 90 |
| Accuracy | 1 | 1 | 0.95 |
| AUC | 1 | 1 | 0.9 |

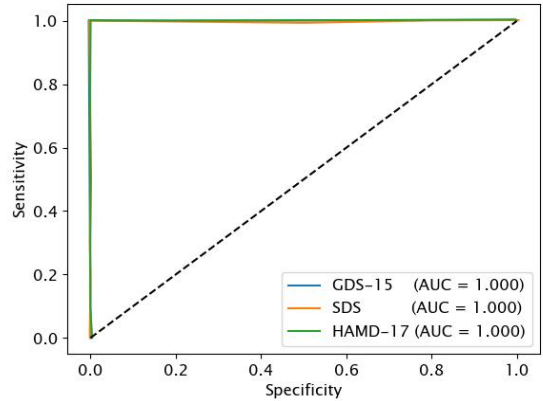

Figure 1: Receiver Operating Characteristic (ROC) curve of the performance of GPT as normal individuals and depressed patients when completing the three selected depression assessment tools.

## 3.2 Role Performance of GPT in Normal Individuals and Patients with Mild, Moderate, and Severe Depression

The experimental results of GPT's performance in role-playing normal individuals and patients with different severity levels of depression were evaluated using the three selected scales for depression assessment (HAMD-17, SDS, and GDS-15), as shown in Table 2 and Figure 2. It can be seen that in the experiment where GPT played the role of a normal individual, the generated responses were basically consistent with the normal individual rating criteria based on the three selected depression assessment tools (HAMD-17, SDS, and GDS-15). However, for the role-playing effect on patients with different levels of depression, the performance of GPT decreased. Specifically, the responses generated by GPT when playing the role of patients with severe depression were more consistent with the rating standards for severe depression levels on all three assessment tools. However, the responses generated by GPT when playing the role of patients with mild and moderate depression were not consistent with the corresponding rating standards on their corresponding assessment tools.

Table 2: Experimental Results of the Performance of GPT as Normal Individuals and Depressed Patients with Different Levels of Depression Severity When Completing the Three Selected Depression Assessment Tools

| Index | GDS-15 | SDS | HAMD-17 |
|---|---|---|---|
| Normal | | | |
| sensitivity(%) | 100 | 90 | 90 |
| specificity(%) | 100 | 100 | 100 |
| Accuracy | 1 | 0.97 | 0.97 |
| AUC | 1 | 1 | 1 |
| MD | | | |
| sensitivity(%) | 0 | 50 | 20 |
| specificity(%) | 100 | 97 | 93 |
| Accuracy | 0.75 | 0.85 | 0.75 |

|      | AUC | 0.51 | 0.66 | 0.57 |
|------|-----|------|------|------|
| MOD  | sensitivity(%) | 0 | 50 | 50 |
|      | specificity(%) | 100 | 83 | 73 |
|      | Accuracy | 0.75 | 0.75 | 0.68 |
|      | AUC | 0.71 | 0.67 | 0.62 |
| SD   | sensitivity(%) | 100 | 100 | 90 |
|      | specificity(%) | 33 | 83 | 83 |
|      | Accuracy | 0.5 | 0.88 | 0.85 |
|      | AUC | 0.8 | 1 | 0.95 |

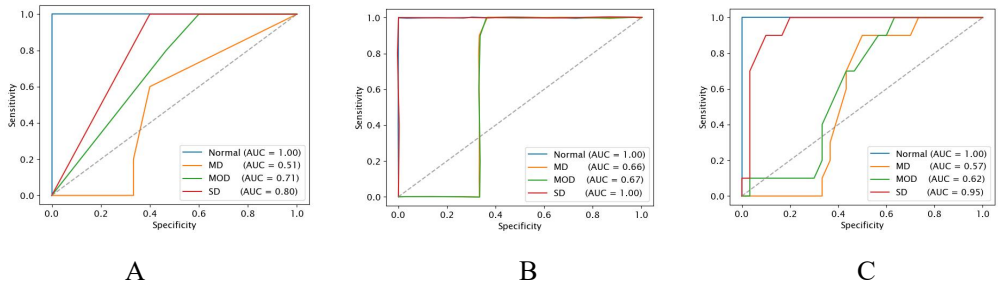

A         B         C

Figure 2: Receiver Operating Characteristic (ROC) curve of the performance of GPT as normal individuals and depressed patients at different levels of depression severity when completing the three selected depression assessment tools. Normal - normal individuals; MD - mild depression; MOD - moderate depression; SD - severe depression. A. GDS-15 B. SDS C. HAMD-17.

### 3.3 Comparison of Test Results from Scales with Different Sensitivities

This study compared the evaluation results obtained by GPT for scales with different sensitivities. According to the experimental performance, when GPT played the roles of either a normal individual or a depressed patient in the selected assessment tools, the generated responses were basically consistent with the expected results. However, when GPT played the roles of patients with different degrees of depression on the selected assessment tools, there were differences in the generated responses. Among them, when GPT played the roles of patients with mild, moderate, or severe depression to complete the GDS-15, the generated responses were consistent with the rating standard for severe depression level, and the sensitivity was 0. This may be due to the fact that the assessment tool only requires binary responses (yes/no) and lacks the factor of multi-level sensitivity. In contrast, when GPT played the roles of patients with mild, moderate, or severe depression on the HAMD-17 and SDS assessment tools, both tools used multi-level rating scales, and the generated responses had a noticeable hierarchical structure and were more consistent with their corresponding rating criteria.

### IV. Discussion：

**4.1** GPT accurately simulates the thinking patterns of both healthy individuals and patients with depression when undertaking depression-related assessments. This study found that GPT can accurately simulate the thinking and behavior of both healthy individuals and patients with depression when completing selected depression assessment scales. GPT technology has great potential in accurately understanding and reflecting depressive symptoms. Previous depression-related studies were limited by the number of participants as they required recruiting subjects for research. However, this study found that GPT technology can accurately simulate the thinking and behavior of both healthy individuals and patients with depression when completing selected depression assessment scales, thus potentially overcoming the limitation of subject numbers in depression-related research and offering a new possibility. Additionally, this research suggests that GPT has significant application potential in the assessment and treatment of depression, providing useful support for clinical doctors and patients.

**4.2** The simulation accuracy of GPT in simulating the thinking patterns of individuals with different degrees of depression during depression-related assessments still needs improvement. This study found that there is a certain degree of simulation bias in GPT when simulating individuals with different degrees of depression. GPT has difficulty understanding the psychological state of individuals with mild and moderate depression, with its simulation results tending to lean towards more severe depressive symptoms. This may be related to the different emotional, behavioral, and thinking patterns exhibited by individuals with different degrees of depression. Individuals with mild and moderate depression possess certain skills and complexity in recognizing depression symptoms, resulting in limitations for GPT in understanding and accurately reflecting their psychological states.

**4.3** GPT's simulation of individuals with different depressive symptoms is more effective on scales with higher sensitivity. This study found that the simulation of individuals with different depressive symptoms by GPT is influenced by the depression scales used. Compared to the use of scales with lower sensitivity, the use of scales with higher sensitivity can improve the accuracy and consistency of GPT's simulation of depressive individuals. This may imply that scales with higher sensitivity appropriately evaluate depressive symptoms, thereby assisting GPT in understanding and simulating the inner states of individuals with depression more accurately. Furthermore, previous scale designs were typically based on subjective experiences of doctors, while GPT may offer new possibilities for scale research and design. GPT may play an important role in developing more effective depression scales and better assessing depressive symptoms.

In addition, the depression scale selected in this study, GDS-15, is primarily used for elderly individuals, which may have an impact on the research results, but it is not discussed in this paper.

**V. Conclusion：**

This study highlights the potential of GPT technology in accurately understanding and reflecting depressive symptoms. It has the ability to simulate the thinking and behavior of both healthy individuals and patients with depression when completing selected depression assessment scales. This offers a new possibility for overcoming the limitation of subject numbers in depression-related research. However, the simulation accuracy of GPT in individuals with different degrees of depression still needs improvement, particularly in understanding individuals with mild and moderate depression. The use of scales with higher sensitivity can enhance the accuracy and consistency of GPT's simulation. Moreover, GPT may play a vital role in developing more effective depression scales and better assessing depressive symptoms. Future research should further explore the impact of using scales designed for specific populations to ensure comprehensive results.

**Reference：**